\documentclass[prd,twocolumn,tightenlines,showpacs,nofootinbib,amsfonts,amssymb,amsmath]{revtex4-1}
\usepackage{graphicx}
\begin{document}
\newcommand{\etal}{{\it et al.}}
\newcommand{\bx}{{\bf x}}
\newcommand{\bn}{{\bf n}}
\newcommand{\bk}{{\bf k}}
\newcommand{\dd}{{\rm d}}
\newcommand{\dslash}{D\!\!\!\!/}
\def\ga{\mathrel{\raise.3ex\hbox{$>$\kern-.75em\lower1ex\hbox{$\sim$}}}}
\def\la{\mathrel{\raise.3ex\hbox{$<$\kern-.75em\lower1ex\hbox{$\sim$}}}}
\def\beq{\begin{equation}}
\def\eeq{\end{equation}}

\leftline{UMN-TH-3129/12}

\vskip-2cm
\title{Stability analysis of chromo-natural inflation and  possible evasion of Lyth's bound}

\author{Emanuela Dimastrogiovanni$^1$, Marco Peloso$^{1}$}
\affiliation{
$^1$School of Physics and Astronomy, University of Minnesota, Minneapolis, 55455, USA
}

\vspace*{2cm}

\begin{abstract} 
We perform the complete stability study of the model of chromo-natural inflation (Adshead and 
Wyman '12), where, due to its coupling to a SU(2) vector, a pseudo-scalar inflaton $\chi$ slowly rolls on a 
 steep potential. As a typical example, one can consider  an axion with a sub-Planckian decay constant $f$. 
 The   model was recently studied (Dimastrogiovanni, Fasiello, and Tolley '12) in the 
 $m_g >> H$ limit, where  $m_g$ is the mass of the fluctuations of the vector field, and $H$ the Hubble rate.  We show that the inflationary solution is stable  for $ m_g > 2 H$, while it  otherwise experiences a  strong instability due to scalar perturbations in the sub-horizon regime.  The tensor perturbations  are instead enhanced at large $m_g$, while  the vector ones remain perturbatively small.  Depending 
on the parameters, this model can give a chiral gravity wave signal that can be detected in ongoing or forthcoming CMB experiments.
This detection can occur even if, during inflation, the inflaton spans an interval of size  $\Delta \chi = {\rm O  } \left( f \right)$ which is 
some orders of magnitude below the Planck scale, evading a well known bound that holds for a free inflaton (Lyth '97).
The spectral tilt of the scalar perturbations typically decreases with decreasing $m_g$. Therefore the simultaneous requirements of stability,  sufficiently small tensor-to-scalar ratio, and  sufficiently flat scalar spectrum  can pose nontrivial bounds on the parameters of the model.
\end{abstract}

\date{\today}

\maketitle

\section{Introduction} 
\label{sec:introduction}

Inflation is a successful paradigm for the physics of the early Universe \cite{Linde:2005ht}. Besides solving the classical problems of modern cosmology  (e.g. the flatness, entropy, and horizon problems), it provides primordial perturbations in perfect agreement  with the observations   \cite{Komatsu:2010fb,Hinshaw:2012fq}. A challenge for inflationary models is to protect the required flatness of the inflaton potential against radiative corrections. This protection can be provided by an approximate shift symmetry as in models of natural inflation \cite{Freese:1990rb,Freese:2004un}. The symmetry can be broken by a controllably small amount, the most known example of this is the case of an axion field acquiring a potential from instantons. The application of this to inflation, however, requires a greater than Planckian axion decay constant $f$ \cite{Savage:2006tr}, which may not be stable against gravitational corrections \cite{Kallosh:1995hi} and which may be impossible to realize in string theory  \cite{Banks:2003sx}. Proposed solutions to this problem include  using two \cite{sol1} or more \cite{sol2} axions, which provide an effective large scale evolution even if the decay constants of the original axions are sub-Planckian,  requiring nontrivial compactifications in string theory  \cite{sol3}, suitably coupling the axion to a $4$-form \cite{sol4}, modifying the axion kinetic term \cite{sol5}, and slowing down the axion evolution through particle production \cite{sol6a,sol6b} as in warm inflation \cite{Berera:1995ie}.

In particular, in the mechanism of \cite{sol6a} the dissipation occurs through the production of a U(1) field coupled to the inflaton $\chi$ through the interaction $\chi F {\tilde F}$ (where $F$ is the U(1) field strength, and ${\tilde F}$ its dual). Ref.  \cite{Adshead:2012kp} 
 showed that this coupling can also affect the background evolution  if the U(1) field is replaced by a SU(2) field with 
a nonvanishing vacuum expectation value (vev). Specifically, due to the interaction with the vev of the vector multiplet, the inflaton can be in slow roll even if its potential would otherwise (i.e., in absence of this interaction) be too steep to give inflation.~\footnote{This mechanism has been extended to a Chern-Simons interaction in \cite{Martinec:2012bv}. For  recent reviews of vector fields in inflation, see 
\cite{reviews}.}
Such a model has been dubbed \textit{chromo-natural inflation} in  \cite{Adshead:2012kp}.
~\footnote{Chromo-natural inflation has trajectories in common with  the so called   \textit{gauge-flation} model  \cite{Maleknejad:2011jw,Maleknejad:2011sq} - a model  characterized by a SU(2) field with  a $(F_{\mu\nu}\tilde{F}^{\mu\nu})^{2}$ term besides the usual 
$F_{\mu\nu}F^{\mu\nu}$ Yang-Mills term - in the limit in which  the axion is  close to the bottom of its potential \cite{Adshead:2012qe,SheikhJabbari:2012qf}. Perturbations of gauge-flation were studied in the last of \cite{reviews}.
}

 The model is characterized by the action
 \begin{eqnarray}
S&=&\int  d^{4}x\sqrt{-g}\Big[\frac{M_{P}}{2}R-\frac{1}{4}F_{\mu\nu}^{a}F^{a \, \mu\nu}-\frac{1}{2}\left(\partial\chi\right)^{2}\nonumber\\&-&V\left(\chi\right)+\frac{\lambda}{8f\sqrt{-g}}\chi\epsilon^{\mu\nu\rho\sigma}F_{\mu\nu}^{a}F_{\rho\sigma}^{a}\Big],
\label{action}
\end{eqnarray}
where $\chi$ is the axion inflaton and $F_{\mu\nu}^{a}=\partial_{\mu}A_{\nu}^{a}-\partial_{\nu}A_{\mu}^{a}-g\epsilon^{abc}A_{\mu}^{b}A_{\nu}^{c}$. We use the convention   $\epsilon^{0123}=1$ for the Levi-Civita tensor.  The vector field has the vev \footnote{We follow the standard convention of using greek letters for space-time indices, $i,j,k \dots$ for space indices, and $a,b,c,\dots$ for internal SU(2) indices. The index $a$ should not be confused with the scale factor $a \left( t \right)$, that enters in the line element as  $ds^{2}=-dt^{2}+a^{2}(t)\delta_{ij}dx^{i}dx^{j}= a^2 \left( \tau \right) \left[ - d \tau^2 + \delta_{ij}dx^{i}dx^{j} \right]$. We denote by dot a derivative with respect to physical time $t$, and by prime a  derivative with respect to conformal time $\tau$.} 
\begin{eqnarray}
A_{0}^{a}=0,\quad\quad A_{i}^{a}=\delta^{a}_{i}a(t)Q(t).
\label{A-vev}
\end{eqnarray}
which is chosen to give isotropic expansion. We note that, for a generic theory with vector fields, an isotropic solution may be unstable against anisotropic perturbations. For instance, this can be the case  when a dilaton-like coupling $f \left( \varphi \right) F^2$ of a scalar inflaton $\varphi$ to a SU(2) field with the vev (\ref{A-vev}) or to an orthogonal U(1) triplet is arranged so to produce a scale invariant spectrum for the vectors.~\footnote{This effect originates from the sum of the IR modes that, in general, strongly breaks isotropy \cite{Bartolo:2012sd}. It is however possible that we live in a realization of inflation where this effect is small, as we believe that must be assumed in the computations of  \cite{hair}.  Analogous considerations may apply to the model of \cite{hair2}. } Isotropy is instead preserved if the vector fields are massive \cite{isotropy-vectors}, as in the current model (if the mass arises from an explicit breaking, one should also check that  the theory has no ghosts  \cite{ghosts}).

Ref.  \cite{Adshead:2012kp} performs a thorough analysis of the background evolution of the model. The study of the chromo-natural inflation theory at the perturbative level was recently performed in \cite{Dimastrogiovanni:2012st}, for
\begin{equation}
m_g^2 \equiv 2 g^2 Q^2 \gg H^2 \,\,,
\label{mg}
\end{equation}
where $m_g$ is the mass of the vector field fluctuations in this limit \cite{Dimastrogiovanni:2012st} (as we also discuss below). When this condition is realized, the vector field can be integrated out while, at the same time, leaving its mark on the inflationary dynamics (this is an explicit realization of the so called \textit{gelaton} \cite{Tolley:2009fg} mechanism). In this limit, chromo-natural inflation is equivalent to a single scalar field $P(X,\chi)=X+X^{2}/\Lambda^{4}-V(\chi)$ theory (where $X\equiv-\left(\partial\chi\right)^{2}/2$ and $\Lambda\equiv 8f^{4}g^{2}/\lambda^{4}$). The non-canonical kinetic term precisely encodes the effect of the gauge fields. 

In \cite{Dimastrogiovanni:2012st} it is also shown that an effective field theory equivalence of chromo-natural inflation to a non-canonical $P(X,\chi)$ Lagrangian holds as long as $m_g^2 > 8 H^2$. Beyond this limit, a general perturbative study of the dynamics of the gauge field and the axion becomes necessary in order to test the stability of the theory and formulate its predictions. This is precisely the scope of the present work. We perform a full linear order study of scalar, tensor and vector perturbations and we show that the inflationary background solution of the model is stable  for $m_g > 2 H$, and it is otherwise unstable.

A full phenomenological study of the model is beyond our purposes. Nonetheless, we explore the scalar and tensor modes production
for a given choice of $f \ll M_p$ and $\lambda \gg 1$ for which the coupling with the vector fields is crucial to ensure slow-roll. Specifically, we choose $f = 10^{-2} M_p$, and $\lambda = 500$, while $m_g/H$ is a free parameter controlled by the value of $g$. The stability condition $m_g/H >2$ provides a lower bound for this ratio. Too small values are also excluded because they lead to a too red spectrum of the scalar modes (we obtained this numerically; this behavior is also seen in the analytic study of  \cite{Dimastrogiovanni:2012st}  in the regime of validity of their analysis). On the  other hand,  the amount of gravity wave signal increases with  $m_g/H$,  and a  level which can be observed in the current or forthcoming experiments \cite{Baumann:2008aq} can be obtained even if the inflaton in the model spans a range of some orders of magnitude below $M_p$, evading the so called Lyth bound \cite{Lyth:1996im} (in contrast to what was expected in \cite{Adshead:2012kp}). Therefore, requiring a stable solution, with a sufficiently flat scalar power spectrum, and sufficiently small tensor modes provides constraints on the parameters of the model that go in opposite directions. For our choice  $f = 10^{-2} M_p$, and $\lambda = 500$ we could not find any acceptable solution. We expect that the situation should improve at larger $f$, where the  inflaton potential becomes flatter (and the model approaches  conventional slow roll inflation).

The paper is organized as follows: in Sec.~\ref{sec:model} we review the background evolution of the model for different (stable or unstable) ranges of the theory; in Sec.~\ref{sec:eqs-perturbations} we start the perturbation analysis by introducing the most general decomposition of the metric, the  inflaton,  and the gauge field fluctuations, identifying the physical degrees of freedom and verifying that scalar, vector and tensor perturbations decouple at linear order; we also quantize the system and define the cosmological correlators; in Secs.~\ref{sec:tensors}, \ref{sec:vectors} and \ref{sec:scalars} we study the fluctuations for, respectively, tensor, vector and scalar modes; finally in Sec.~\ref{sec:conclusions} we draw our conclusions. \\We supplement our perturbation analysis in Appendix~\ref{app-dg-scalar}, where we show that neglecting scalar metric fluctuations does not affect the stability analysis of the model (while it considerably simplifies the computations).

\section{The model and the background solution}
\label{sec:model}

Chromo-natural inflation is described by the action (\ref{action}) of a pseudo-scalar field $\chi$ coupled to an $SU(2)$ gauge multiplet.
The vector multiplet has the vev (\ref{A-vev}).  The \textit{00} component of Einstein equations reads
\begin{eqnarray}
\label{zerozero}
3 M_{P}^2 H^2 =\frac{3}{2}\left( \dot{Q} + H Q \right)^2+\frac{3}{2}g^{2}Q^{4}+\frac{\dot{\chi}^{2}}{2}+V\left(\chi\right).
\end{eqnarray}
Inflation can only occur if the  potential $V \left( \chi \right)$ is the major contribution to the total energy density, i.e. $3H^{2}M_{P}^{2} \simeq V$; the parameters $\epsilon_{1}\equiv Q^{2}/ M_{P}^{2}$, $\epsilon_{2}\equiv g^{2}Q^{4}/ H^{2}M_{P}^{2}$, $\epsilon_{Q}\equiv\dot{Q}^{2}/ H^{2}M_{P}^{2}$ and $\epsilon_{\chi}\equiv\dot{\chi}^{2}/2 H^{2}M_{P}^{2}$ are all much smaller than unity.\\

The equations of motion for the inflaton and the gauge field are: 
\begin{eqnarray}\label{eqone}
&&
\ddot{\chi}+3H\dot{\chi}-\frac{\mu^{4}}{f}\sin\left(\frac{\chi}{f}\right)=-\frac{3g\lambda}{f}Q^2\left(\dot{Q}+HQ\right),\\\label{eqtwo}
&&
\ddot{Q}+3H\dot{Q}+\left(\dot{H}+2H^{2}\right)Q+2g^2Q^3=\frac{g\lambda}{f}Q^{2}\dot{\chi}.
\end{eqnarray}
If we neglect $\ddot{\chi}$, $\ddot{Q}$ and $\dot{H}$, one can solve Eqs.~(\ref{eqone}-\ref{eqtwo}) for $\dot{\chi}$ and $\dot{Q}$ \cite{Adshead:2012kp}
\begin{eqnarray}
&&
\dot{\chi}\simeq \frac{g\lambda f Q^{2} H\left(\frac{2g^{2}Q^{3}}{H}-HQ-\frac{fV_{,\chi}}{g\lambda Q^{2}}\right)}{3f^{2}H^{2}+g^{2}\lambda^{2}Q^{4}}, \\
&&
\dot{Q}\simeq-\frac{HQ\left(2f^{2}H^{2}+2g^{2}f^{2}Q^{2}+g^{2}\lambda^{2}Q^{4}\right)+\frac{g\lambda Q^{2}f V_{,\chi}}{3}}{3f^{2}H^{2}+g^{2}\lambda^{2}Q^{4}}.\nonumber\\
\end{eqnarray}
In these equations we then assume
\begin{eqnarray}
3f^{2}H^{2}\ll g^{2}\lambda^{2}Q^{4},\quad\quad \lambda^{2}Q^{2}\gg 2 f^{2} \,,
\end{eqnarray}
and we obtain 
\begin{eqnarray}\label{eqfour}
&&
\dot{\chi}\simeq \frac{f H}{g \lambda Q^{2}}\left(\frac{2g^{2}Q^{3}}{H}-HQ-\frac{f V_{,\chi}}{g\lambda Q^{2}}\right),\\\label{eqthree}
&&
\dot{Q}\simeq -HQ- \frac{f V_{,\chi}}{3g\lambda Q^{2}}.
\end{eqnarray}
Eq.~(\ref{eqthree}) can be rewritten as an equation of motion for the gauge field in terms of an effective potential
\begin{eqnarray}
H\dot{Q}+\frac{\partial V_{\rm eff}\left(Q \right)}{\partial Q}=0,\quad V_{\rm eff}\equiv \frac{H^{2}Q^{2}}{2}-\frac{f H V_{\chi}}{3g\lambda Q}
\end{eqnarray}
where $V_{\rm eff}$ is minimized by 
\begin{eqnarray}
Q_{\rm min}= \left(\frac{\mu^{4}\sin(\chi/f)}{3g\lambda H}\right)^{1/3}.
\label{Qmin}
\end{eqnarray}

In the minimum, the two terms of $V_{\rm eff}$ are parametrically equal to each other.  Therefore, the same is true for the last two terms 
in (\ref{eqfour}). As a consequence, the first term on the right hand side of  (\ref{eqfour}) dominates over the other two in the regime $m_g^2 \gg H^2$ studied in  \cite{Dimastrogiovanni:2012st}. This implies that  \cite{Dimastrogiovanni:2012st}
\begin{equation} 
m_g^2 = 2 \,  g^2 Q^2 \gg H^2 \;\;\Rightarrow\;\; m_g^2 \simeq g \lambda Q \frac{\dot{\chi}}{f}
\label{mg-Qchidot}
\end{equation}

For general values of the parameters, inserting $Q_{\rm min}$, Eq.~(\ref{eqfour}) becomes
\begin{eqnarray}
\label{chid}
\dot{\chi}\simeq \frac{2}{3^{2/3}}\frac{f^{4/3}}{\lambda^{4/3}}\frac{3\frac{\lambda^{2/3}}{f^{2/3}}H^{8/3}+3^{1/3}g^{4/3}\left(-V_{,\chi}\right)^{2/3}}{g^{2/3}H^{1/3}\left(-V_{,\chi}\right)^{1/3}}.
\end{eqnarray}
By looking at Eq.~(\ref{eqone}), one realizes that, when the gauge field settles in its minimum, it originates an actual damping term for the motion of the axion. Notice also that, when $Q=Q_{min}$, the kinetic energy of the fields can be disregarded (see e.g. Eq.~(\ref{eqthree})); therefore the slow-roll parameter (which would normally receive contributions also from $\epsilon_{Q}$ and $\epsilon_{\chi}$) will be, instead, mostly due to $\epsilon_{1}$ and $\epsilon_{2}$
\begin{eqnarray}
\label{slowrolld}
\epsilon\equiv-\frac{\dot{H}}{H^{2}}\simeq\frac{Q^{2}}{M_{P}^{2}}+\frac{g^{2}Q^{4}}{H^{2}M_{P}^{2}}.
\end{eqnarray}
Finally we have
\begin{eqnarray}
\eta\equiv\frac{\dot{\epsilon}}{H\epsilon}\simeq \frac{2g^{2}Q^{4}}{H^{2}M_{P}^{2}}+\frac{\dot{Q}}{HM_{P}^{2}\epsilon}\left(2Q+\frac{4g^{2}Q^{3}}{H^{2}}\right).
\end{eqnarray}
From Eq.~(\ref{chid}) we can compute the number of e-foldings
\begin{eqnarray}
\label{efoldings}
N\simeq \int dx \frac{\frac{3^{1/3}}{2}\left(\frac{\mu}{M_{P}}\right)^{4/3}g^{2/3}\lambda^{4/3}\left(\sin x\right)^{1/3}\left(1+\cos x\right)^{2/3}}{\left(\frac{\mu}{M_{P}}\right)^{8/3}\lambda^{2/3}\left(1+\cos x\right)^{4/3}+3^{2/3}g^{4/3}\left(\sin x\right)^{2/3}} \; , \nonumber\\
\label{efolds}
\end{eqnarray}
where we defined $x\equiv\chi/f$.\\
Let us introduce the parameter $y$ 
\begin{equation}
\label{y}
y \equiv \left( \frac{\lambda \mu^4}{3 g^2 M_p^4} \right)^{2/3};
\end{equation}
in terms of which the expression (\ref{efolds}) rewrites
\begin{equation}
N \simeq \frac{3 \sqrt{y} \lambda}{2} \int   \frac{dz \, z}{2 y + z - y z^3} \;\;,\;\; z \equiv \left( 1 - \cos \, \frac{\chi}{f} \right)^{1/3}
\label{efolds2}
\end{equation}
An upper bound on $N$ can be obtained by ``pretending'' that the slow roll approximation holds at all values of $\chi$, corresponding to 
$0 \leq z \leq 2^{1/3}$. The resulting upper limit is maximized for $y \simeq 1$, where it evaluates to $N \la 0.6 \lambda$. In the $y \ll 1$ limit, the upper limit acquires the simple form $N < \frac{3 \sqrt{y} \, \lambda}{2^{2/3}}$.

In terms of $y$, the slow roll solutions read
\begin{eqnarray}
&& H \simeq \frac{\mu^2}{\sqrt{3} M_p} \sqrt{1+\cos x} 
\equiv  \frac{\mu^2}{\sqrt{3} M_p} f_H \left( x \right),
 \nonumber\\
&& Q \simeq M_p \frac{y^{1/4}}{\lambda^{1/2}} \left( \frac{\sin x}{\sqrt{1+\cos x}} \right)^{1/3} 
\equiv \frac{M_p}{\lambda^{1/2}} f_H \left( x \right) f_A \left( x \right), \nonumber\\
&& \dot{\chi} \simeq   \frac{2 f \mu^2}{\sqrt{3} M_p \lambda} \, \frac{y \left( 1 + \cos x \right)^{4/3} + \left(  \sin x \right)^{2/3}}{\sqrt{y}
\left( 1 + \cos x \right)^{1/6} \left( \sin x \right)^{1/3} }
\simeq \frac{H f}{\lambda} f_{\dot{\chi}} \left( x \right), \nonumber\\
\label{slow}
\end{eqnarray}
where we have defined
\begin{eqnarray}
&&
f_H \equiv  \sqrt{1+\cos x} 
\;,\; f_A \equiv \frac{y^{1/4} \left( \sin x \right)^{1/3}}{\left( 1 + \cos x \right)^{2/3}}, \nonumber\\
&& f_{\dot{\chi}} \equiv 2 \,  \frac{y \left( 1 + \cos x \right)^{4/3} + \left(  \sin x \right)^{2/3}}{\sqrt{y}
\left( 1 + \cos x \right)^{2/3} \left( \sin x \right)^{1/3} }.
\end{eqnarray}

\begin{figure}[ht!]
\centerline{
\includegraphics[width=0.4\textwidth,angle=-90]{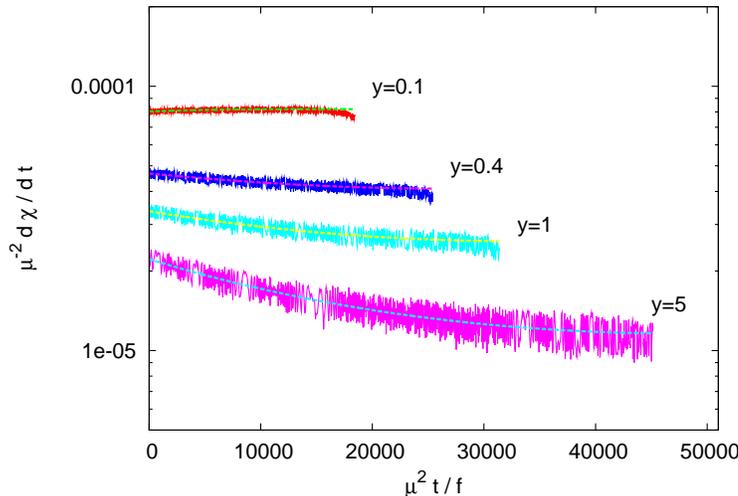}
}
\caption{
Evolution of $\dot{\chi}$ during inflation for four different values of $y$. The other parameters are $\lambda = 500$ and $f=10^{-2} M_p$.
The evolutions shown correspond to $60$ e-folds of inflation.
}
\label{fig:bck}
\end{figure}

Figure \ref{fig:bck} allows to appreciate the accuracy of these slow roll approximations. In the figure we show the evolution of $\dot{\chi}$ for four choices of $y$, comparing the exact evolution with the slow roll approximation (\ref{slow}). 
 \footnote{The initial conditions for the numerical evolution shown in the Figure  are chosen as follows: we fix an initial value for $\chi$; we then employ (\ref{slow}) to have the initial value for $\dot{\chi}$ and $Q$. We set $\dot{Q} = 0$, and we then obtain the initial value for $\dot{a}$ by solving the $00$ Einstein equations exactly at the initial time. We make sure that the initial value of $\chi$ leads to more than  $60$ e-folds of inflation (so that the slow-roll solution can be achieved; we note that the system starts slightly displaced from the slow roll solution, since we set $\dot{Q} = 0$; the displacement is however very small, since $\dot{Q} \ll H Q$ in the slow roll solution, and 
 the background evolution quickly approaches the slow roll solution). The evolutions shown in Figure \ref{fig:bck} are restricted to the final $60$ e-folds of inflation.} 
In all cases, we take $f=0.01 M_p$, as the main purpose of this model is to allow slow roll for $f \ll M_p$. This is ``compensated'' by $\lambda \gg 1$, and in all cases we fixed for definiteness $\lambda = 500$. We note that fixing the value of $y$ does specify a relation between $\mu$ and $g$, but does not fix these two values. As typical in inflationary models, the parameters can be specified only from  fixing the normalization of the power spectrum of the scalar modes to the observed value $P_\zeta \simeq 2.5 \cdot 10^{-9} $ \cite{Komatsu:2010fb}. The evolutions shown in the figure cover $60$ e-folds of inflation. 

We note from the figure that $\dot{\chi}$ performs very small oscillations around the slow roll solution. We believe that they are due to the fact that $Q \left( t \right)$ is tracking a time dependent minimum (\ref{Qmin}) (we do not show this here, but also this tracking is extremely accurate during the slow roll phase). The oscillations appear somewhat large in the logarithmic scale chosen, but we see that  they do not lead to any net departure from the slow roll solution. Moreover, in Section \ref{sec:scalars} we compare full numerical solutions of the scalar perturbations, obtained using the full numerical background solutions, with analytical solutions, for which the slow roll approximations are used, and we also find excellent agreement. As we discussed after eq. (\ref{efolds2}), decreasing $y$ in the $y<1$ region, while keeping the other parameters (including the initial value of $\chi$) fixed, decreases the amount of inflation. As a consequence,    $ y  \ll 1$ can result in sufficient inflation only if  the inflaton is initially  close to the top of the potential. Another way to express this is to note that the inflaton rolls faster as $y$ decreases (with the other parameters kept fixed), as it is clear both from the slow roll expression (\ref{slow}) and from the figure.

Perhaps the most surprising feature of the model is that, although all the solutions shown in Figure \ref{fig:bck} appear to be acceptable inflationary solutions (in all cases the slow roll solution appears to be an attractor; we note however that the background dynamics only probes homogeneous departures from the slow roll solution), and although they only differ from each other by the value of $y$, the background solution with $y=5$ is unstable, while the other ones are stable. This emerges from the study of the scalar perturbations around these solutions that we perform below.

\section{Linearized equations for the perturbations}
\label{sec:eqs-perturbations}

In this Section we discuss at the formal level how we compute, we quantize, and we solve the linearized theory for the perturbations of the model around the background solution discussed in the previous Section.  The discussion is divided in two Subsections. In the first Subsection we discuss how the perturbations can be divided into three groups, decoupled from each other at the linearized level. In the second Subsection we give the form of the quadratic action and we discuss how we compute the corresponding linearized theory for the perturbations.

\subsection{Decomposition}
\label{subsec:deco} 

There are $23$ perturbations in the system, one of the inflaton, $12$ of the SU(2) vector field, and $10$ of the metric, that we decompose as: 
\begin{eqnarray}
\chi & = & \chi + \delta \chi \nonumber\\
A_0^a & = & a \left( Y_a + \partial_a Y \right) \nonumber\\
A_i^a & = & a \Big[ \left( Q + \delta Q \right) \delta_{ai} + \partial_i \left( M_a + \partial_a M \right) \nonumber\\
& & \quad\quad\quad\quad + \epsilon_{iac} \left( U_c + \partial_c U \right) + t_{ia} \Big] \nonumber\\
g_{00} & = & - a^2 \left( 1 - 2 \phi \right) \nonumber\\
g_{0i} & = & a^2 \left( B_i + \partial_i B \right) \nonumber\\
g_{ij} & = & a^2 \left[ \left( 1 + 2 \psi \right) \delta_{ij} + 2 \partial_i \partial_j E + \partial_i E_j + \partial_j E_i + h_{ij} \right] \nonumber\\
\label{deco1}
\end{eqnarray}
In this expression, $a=1,2,3$ is the SU(2) index (we also denote by $a$ the scale factor, as there is no ambiguity between it and an SU(2) index), while $i=1,2,3$ ranges over the spatial coordinates. We denote as ``tensor modes'' the perturbations $t_{ia}$ and $h_{ij}$, which we impose to be transverse ($\partial_i h_{ij} = \partial_i t_{ia} = \partial_a t_{ia} = 0$) and traceless ($t_{ii} = h_{ii} = 0$); due to these properties, the tensor sector contains $4$ perturbations. We denote as ``vector modes'' the perturbations $Y_a,M_a,U_c,B_i,E_i$, which we impose to be transverse ($\partial_i Y_i = \dots = \partial_i E_i = 0$); due to this, the vector sector contains $10$ perturbations. We denote as ``scalar modes'' the remaining $9$ perturbations. 

We point out that the terms  ``tensor/vector/scalar" are appropriate for the  perturbations of the metric and of the inflaton, as they indicate how these modes transform under a spatial rotation. We extend this terminology also to the perturbations of the vector field, following the notation of  \cite{Maleknejad:2011sq}, even if, strictly speaking, these terms are inappropriate (given that the SU(2) index has been used in the decomposition). We nonetheless adopt it since the fact that   the vector vev is diagonal ($\langle A^a_i \rangle \propto \delta^a_i$) plus the transversality properties that we have imposed guarantee that the tensor / vector / sectors that we have defined above remain decoupled from each other at the linearized level \cite{Maleknejad:2011sq}. 

We Fourier transform these modes
\begin{equation}
\delta \left( t ,\, {\bf x} \right) = \int \frac{d^3 k}{\left( 2 \pi \right)^{3/2}} {\rm e}^{i {\bf k} \cdot {\bf x}} \, \delta \left( t , {\bf k} \right)
\end{equation}
where $\delta$ denotes any of the perturbations, and we study the theory in Fourier space. In our stability study, we solve
for the perturbations at the linearized level, and therefore we study a mode with a given momentum ${\bf k}$ (modes of different momenta are coupled to each other at the nonlinear level). We can actually fix the orientation of ${\bf k}$ along the $z-$axis without loss of generality. Starting from a general direction for ${\bf k}$, we rotate the coordinates so that ${\bf k} = k \, {\hat k}$, where ${\hat k}$ is the unit vector along the $z-$axis, and $k>0$. After a general rotation, $\langle A^a_i \rangle $ is no longer proportional to $\delta_i^a$; however we can re-obtain  $\langle A^a_i \rangle = Q \, \delta_a^i$ through a global SU(2) rotation. Therefore, we can set $k_x = k_y = 0$ without loss of generality. This choice simplifies our algebra. 

We need to remove the redundancy associated to general coordinate and SU(2) transformations. Under an infinitesimal coordinate transformation with parameter $\xi^\mu= \left( \xi^0 ,\, \xi_i + \partial_i \xi \right)$,   
\begin{equation}
\psi \rightarrow \psi - {\cal H} \xi^0 \;,\;
E \rightarrow   E - \xi \;,\;
E_i \rightarrow E_i - \xi_i
\end{equation}
and we remove the freedom of infinitesimal coordinate transformations by setting  $\psi = E = E_i = 0$. Consider instead an SU(2) transformation with infinitesimal parameter $\alpha^a = \epsilon_a + \partial_a \epsilon$ (with $\epsilon_a$ transverse). Under this transformation,
\begin{equation}
U \rightarrow U + g Q \epsilon \;,\;  U_i \rightarrow U_i + g Q 
\end{equation}
and we can fix the SU(2) freedom by setting $U=U_i=0$. Clearly, also  other modes of the metric and of the gauge field change under these transformations, and different gauge choices can be made. Our choices are motivated by the fact that (i) they completely fix the freedom, and (ii) they preserve all the $\delta g_{0\mu}$ and $\delta A_0^a$ modes. These perturbations are nondynamical, as they enter in the quadratic  action of the perturbations  without time derivative, and can be immediately integrated out.

With our gauge choices, and with  $k=k_z$, the decomposition (\ref{deco1}) acquires the explicit form
\begin{eqnarray}
\chi & = & \chi + \delta \chi \nonumber\\
A_\mu^1 & = & a  \left( Y_1 ,\, Q + \delta Q + t_+ ,\,    + t_\times ,\, \partial_z M_1   \right) \nonumber\\
A_\mu^2 & = & a  \left( Y_2 ,\,    t_\times ,\, Q + \delta Q - t_+  ,\, \partial_z M_2   \right) \nonumber\\
A_\mu^3 & = & a  \left( \partial_z Y ,\,  0 ,\, 0 ,\, Q + \delta Q + \partial_z \partial_z M \right)
\label{deco2a}
\end{eqnarray}
and
\begin{eqnarray}
g_{\mu \nu} = a^2 \left( \begin{array}{cccc}
- 1 + 2 \phi & B_1 & B_2 & \partial_z B \\
& 1    + h_+ & h_\times & 0  \\
& & 1    - h_+ & 0  \\
& & & 1        
\end{array} \right)
\label{deco2b}
\end{eqnarray}
We verified explicitly that the scalar modes ($\delta \chi ,\, Y ,\, \delta Q ,\, M   ,\, \phi ,\, B    $), the vector modes ($Y_{1,2} ,\, M_{1,2} ,\,   B_{1,2}  $) and the tensor modes ($t_+ ,\, t_\times ,\, h_+ ,\, h_\times$)  are decoupled from each other at the linearized level. Namely, the quadratic action for the perturbations splits into three decoupled parts
\begin{equation}
S_2 = S_{2,{\rm scalar}} +  S_{2,{\rm vector}} +  S_{2,{\rm tensor}}.  
\end{equation}

\subsection{Quantization of coupled systems and correlators}
\label{subsec:quantization}

We have seen that the total action for the perturbations splits in a sum of three decoupled quadratic actions. Let us denote by ${\cal Y}$ the vector formed by the perturbations in one of these three systems. We can perform a transformation
\begin{equation}
{\cal Y}_i = {\cal M}_{ij} \, \Delta_j
\label{Y-Delta}
\end{equation}
so that the action for the array $\Delta$ is of the type
\begin{equation}
S =  \frac{1}{2} \int d \tau d^3 k \left[ \Delta^{' \dagger} \Delta' +  \Delta^{' \dagger} K \Delta - \Delta^\dagger K \Delta' - \Delta^\dagger \Omega^2 \Delta \right]
\label{formal}
\end{equation}
Hermitianity of the  action implies that $K$ is an anti-Hermitian matrix, and $\Omega^2$ a Hermitian matrix; the matrices obtained in the current model are actually real.  In the following we quantize the system (\ref{formal}), so to obtain the initial conditions for the modes, and we give an expression for the correlators between the modes. This discussion summarizes the one of \cite{Gumrukcuoglu:2010yc}.  We first ``rotate''
\begin{equation}
\psi \equiv R \Delta,
\label{psi-delta}
\end{equation}
where $R$ is a unitary matrix (so that the kinetic term in (\ref{formal}) remains canonical), satisfying
\begin{equation}
R' = R K.
\end{equation}
As $K$ is real, $R$ can also be taken  real, and, therefore, orthogonal. We note that $R$ is not uniquely determined by this condition, and we fix it by setting $R = 1$ at the initial time $\tau_{\rm in}$;  the goal of the present discussion is also to understand when this initial time can be set. As we will see, the explicit solution for $R$ is not needed. In terms of the vector $\psi$, the action becomes
\begin{eqnarray}
S & = & \frac{1}{2} \int d \tau d^3 k \left[ \psi'^\dagger \psi' - \psi^\dagger \, {\tilde \Omega}^2 \, \psi \right], \nonumber\\
 {\tilde \Omega}^2 & \equiv & R \left( \Omega^2 + K^T \, K \right) R^T.
\end{eqnarray}

We then introduce the orthogonal matrix $C$ satisfying
\begin{equation}
C^T \, {\tilde \Omega}^2 C = {\rm diag } \left( \omega_1^2 ,\, \omega_2^2 ,\, \omega_3^2 \right) \equiv \omega^2
\end{equation}
and we decompose
\begin{equation}
\psi_i = C_{ij} \left[ h_{jl} a_l + h_{jl}^* a_l^\dagger \right],
\label{deco-psi}
\end{equation}
where $a_i / a_i^\dagger$ destroys / creates a quantum with the frequency $\omega_i$. These operators satisfy the algebra
\begin{equation}
\left[ a_i \left( \vec{k} \right) ,\, a_j^\dagger \left( \vec{p} \right) \right] = \delta^{(3)} \left( \vec{k} - \vec{p} \right) \, \delta_{ij}.
\end{equation}

For the systems that we study, in the deep sub-horizon regime $ \left( a H \right)_{\rm in} \ll k $
\begin{equation}
{\tilde \Omega}_{\rm in}^2 = \left(  \Omega^2 + K^T \, K \right)_{\rm in} \equiv k^2 \, 1 + a^2 H^2 {\cal C}
\label{Omega}
\end{equation}
with ${\cal C}$ constant at leading order in slow roll. Therefore, provided that ${\tilde \Omega}_{\rm in}^2 \simeq k^2 \, 1
$, we can set the initial conditions in the adiabatic vacuum
\begin{equation}
h  \simeq \frac{{\rm e}^{-i \int^\tau d \tau \omega}}{\sqrt{2 \omega}} \, U =
 \frac{{\rm e}^{-i \int_{\tau_{\rm in}}^\tau d \tau \omega}}{\sqrt{2 \omega}}   
\label{h-in}
\end{equation}
where $U$ is a constant arbitrary orthogonal matrix (at the practical level, Eq. (\ref{h-in}) is an approximate solution of the equations of motion at early times; the closer  ${\tilde \Omega}_{\rm in}^2 $ is to $k^2 1$, the better this approximation is. This determines how early $\tau_{\rm in}$ needs to be taken).  The matrix $U$ is unphysical, as it drops from the equations of motion for the modes, and from the physical correlators (which, as we will see, are given in terms of  $h h^\dagger$). The freedom associated to $U$ is the generalization to $N$ fields of the freedom of changing by a constant phase the wave function in the single field case. In the final expression in (\ref{h-in}) we have used the freedom associated to $U$ to set the wave functions to be real at the initial time.
 
Combining (\ref{psi-delta}) and (\ref{deco-psi}), we have
\begin{equation}
\Delta_i = {\cal D}_{ij} a_j +  {\cal D}_{ij}^* a_j^\dagger \;\;\;,\;\;\; {\cal D} = R^T C h.
\end{equation}

Using the fact that  ${\tilde \Omega}_{\rm in}^2 \simeq k^2 1$ at $\tau_{\rm in}$, so that initially  $C \simeq 1$, we arrive at
\begin{equation}
{\cal D}_{\rm in} = \frac{1}{\sqrt{2 k}} \;\;\;,\;\;\;
{\cal D}_{\rm in}' = -i \sqrt{\frac{k}{2}}.
\label{Delta-in}
\end{equation}
We start from these initial condition and evolve the equations of motion for the modes following from (\ref{formal}) 
\begin{equation}
{\cal D}'' + 2 K {\cal D}' + \left( \Omega^2 + K' \right) {\cal D} = 0.
\label{Delta-eom}
\end{equation}

In this discussion, all the expressions are given in Fourier space. Let us denote by $Y_i$ the original fields in real space,
\begin{equation}
Y_i = \int  \frac{d^3 k}{\left( 2 \pi \right)^{3/2}} {\rm e}^{i {\bf k} \cdot {\bf x}} {\cal Y}_i
\end{equation}

We have the two point correlation functions
\begin{eqnarray}
{\cal C}_{ij} \left( \vec{x} ,\, \vec{y} \right) & = & \frac{1}{2} \left\langle \, Y_i \left( \tau ,\, \vec{x} \right) \,  Y_j \left( \tau ,\, \vec{y} \right) + 
 Y_j \left( \tau ,\, \vec{y} \right) \,  Y_i \left( \tau ,\, \vec{y} \right) \, \right\rangle \nonumber\\
  & \equiv & \int \frac{d k}{k} \, \frac{\sin \left( k r \right)}{k r} \, {\cal P}_{ij} \;\;\;,\;\;\; r \equiv \vert \vec{x} - \vec{y} \vert
  \end{eqnarray} 
where the power spectra are given by
\begin{eqnarray}
{\cal P}_{ij} \left( k \right) & = & \frac{k^3}{2 \pi^2} \, {\rm Re } \left[ \left( {\cal Y} \,{\cal Y}^\dagger \right)_{ij} \right] \nonumber\\
& = &  \frac{k^3}{2 \pi^2} \, {\rm Re } \left[ \left( {\cal M} \, {\cal D} \, {\cal D}^\dagger \, {\cal M}^T \right)_{ij} \right].
\label{power}
\end{eqnarray}
These correlators are the theoretical prediction, to be confronted with the statistical average of the corresponding quantities.
The power spectra (\ref{power}) are the generalization to a system of $N$ fields $Y_i$ of the standard power spectrum of single field inflation.

To summarize, starting from the original fields ${\cal Y}_i$ in momentum space,  we perform (\ref{Y-Delta}) to have a canonical kinetic term for $\Delta_i$. We then decompose this field in terms of annihilation / creation operators of the physical quanta in the system (the particles of frequencies $\omega_i$). We work in terms of the coefficients ${\cal D}_{ij}$ of this decomposition. We can set the initial conditions (\ref{Delta-in}) for these coefficients, provided that the initial time is chosen sufficiently early such that $\Omega^2 +  K^T K \simeq k^2$. Starting from these initial conditions, the coefficients ${\cal D}_{ij}$ evolve according to (\ref{Delta-eom}). We will see that for some choice of parameters some of the perturbations become exponentially large on a timescale $\ll H^{-1}$, signaling an instability of the linearized theory. For values of parameters leading to stable solutions, the coefficients enter in observable quantities through the power spectra (\ref{power}).

\section{Tensor modes}
\label{sec:tensors} 

We introduce the two doublets
\begin{equation}
\Delta_{\rm L} \equiv \frac{\Delta_+ + i \Delta_\times}{\sqrt{2}} \;\;\;,\;\;\; 
\Delta_{\rm R} \equiv \frac{\Delta_+ - i \Delta_\times}{\sqrt{2}} 
\end{equation}
for the left and right helicities,  where
\begin{equation}
\Delta_+ \equiv \left( \begin{array}{c}
\frac{a M_p}{\sqrt{2}} \, h_+ \\
\sqrt{2} a t_+ 
\end{array} \right) \;\;\;,\;\;\;
\Delta_\times \equiv \left( \begin{array}{c}
\frac{a M_p}{\sqrt{2}} \, h_\times \\
\sqrt{2} a t_\times 
\end{array} \right). 
\end{equation}
The action for the tensor modes splits into two separate actions, one for the left and one for the right helicity doublet, which are formally of the  type (\ref{formal}). The action for the left-helicity doublet is characterized by
\begin{eqnarray}
K_{12} & = &  \frac{1}{M_p} \left( Q' + {\cal H} \, Q \right) \simeq  a H \,  {\rm O } \left( \frac{1}{\sqrt{\lambda}} \right) \nonumber\\
\end{eqnarray}
and
\begin{eqnarray}
\Omega_{11}^2 & = &  
k^2 - 2 {\cal H}^2 - \frac{1}{M_p^2}  \left( Q' + {\cal H} \, Q \right)^2 + \frac{3 g^2 a^2 Q^4}{M_p^2} + \frac{\chi^{' 2}}{2 M_p^2} 
\nonumber\\
& \simeq & k^2 +  a^2 H^2 \left[ - 2 + {\rm O } \left( \frac{1}{\lambda} \right) \right] \nonumber\\
\Omega_{12}^2 & = & a k \, \frac{2 g Q^2}{M_p} + \frac{\cal H}{M_p}  \left( Q' + {\cal H} \, Q \right) - \frac{g \lambda Q^2 a \chi'}{f M_p} 
\nonumber\\
&& \simeq a k H  {\rm O } \left( \frac{1}{\sqrt{\lambda}} \right) +
   a^2 H^2  {\rm O } \left( \frac{1}{\sqrt{\lambda}} \right) \nonumber\\ 
\Omega_{22}^2 & = &  k^2 - a k \left( 2 g Q + \frac{\lambda}{f} a \chi' \right) + g \frac{  \lambda }{ f }  Q a \chi'   \nonumber\\ 
&&  \simeq k^2 - a k H \, {\cal A}+ a^2 H^2 {\cal B}
\label{left-action}
\end{eqnarray}
where the first expression for each coefficient is exact, while the second one is obtained from the slow roll approximation (\ref{slow}) (we note that $\dot{Q} \ll H \, Q$). The quantities ${\cal A}$ and ${\cal B}$ are both of ${\rm O } \left( \lambda^0 \right)$.  Using the slow roll result (\ref{eqthree}) with $Q$ given by (\ref{Qmin}), they evaluate to 
\begin{equation}
  {\cal A} \simeq \frac{4 g Q}{H} \left( 1 + \frac{H^2}{2 g^2 Q^2} \right) \;\;\;,\;\;\;
 {\cal B} \simeq \frac{2 g^2 Q^2}{H^2} \left( 1 + \frac{H^2}{ g^2 Q^2} \right) 
\label{AB}
\end{equation}
The action for the right-helicity doublet is related to (\ref{left-action}) by $k \rightarrow - k$, signaling the breaking of parity invariance induced by the evolution of the pseudo-scalar inflaton.

The interactions between the  gravity wave and vector field tensor perturbations are slow-roll suppressed. However, the effective frequency squared $\Omega_{22}^2$ of $t_{L}$ turns negative for an intermediate interval of time next to horizon crossing.
\footnote{At late times, the $k-$dependence  becomes negligible, and  the mass term for the two vector perturbations $t_{L/R}$ 
reproduces the value $m_g$ given in  \cite{Dimastrogiovanni:2012st}.} This leads to a tachyonic growth of $t_L$ in this interval of time, and, correspondingly, to a growth of $h_L$.   The same growth does not occur in the right-helicity sector due to the opposite sign of the linear term in $k$. The situation is analogous to that first studied in \cite{sol6a}, where the interaction $\chi F F~$ between a vector field and a pseudo-scalar rolling inflaton results in a tachyonic growth of the vector modes of a given helicity. 

We note from (\ref{AB}) that the length of the tachyonic region increases with increasing 
\begin{equation}
\frac{ m_g }{ H }  = \frac{\sqrt{2} g Q}{H} \simeq \sqrt{ \frac{2}{y} } \, \frac{\sin^{1/3 } x }{\left( 1 + \cos x \right)^{2/3}}
\label{mgH}
\end{equation}
(the last expression is the slow roll approximation).  This corresponds to a larger tensor mode production with growing $\frac{m_g}{H}$.
This corresponds to decreasing $y$ in the numerical examples that we show below.

\section{Vector modes}
\label{sec:vectors}

The vector modes $Y_1,Y_2,B_1,B_2$ are non-dynamical, and can be integrated out. Namely, they enter in the quadratic action of the perturbations without time derivatives, and therefore their equations of motion are algebraic equations in these variables (recall that we are in momentum space). When we solve these equations, we obtain an expression for the non-dynamical modes in terms of the two dynamical modes $M_1$ and $M_2$. We then insert this expression back in to the quadratic action for the vector modes, and obtain an action for $M_{1,2}$ only. In other words, the non-dynamical modes do not introduce additional degrees of freedom, but are completely determined by the dynamical ones. After integrating out the non dynamical modes, we define
\begin{equation}
M_1 = F_1 V_1 + i F_2 V_2 \;\;,\;\; M_2 = i F_1 V_1 + F_2 V_2
\end{equation}
where
\begin{equation}
F_{1,2} \equiv \frac{\sqrt{M_p^2 \left( \pm k + g a Q \right)^2 + g^2 a^2 Q^2 \left( M_P^2 + 2 Q^2 \right)}}{\sqrt{2} g k M_p a^2 Q} 
\end{equation}
(with the upper $+$ sign corresponding to $F_1$ and the lower $-$ sign corresponding to $F_2$). The modes $V_{1,2}$ are the canonical modes of the system, and are decoupled from each other at the linearized level: %
\begin{eqnarray}
& & 
S_{2,{\rm vector}} = \frac{1}{2} \int d \tau d^3 k \Bigg[ \vert V_+'  \vert^2 - \Omega_{v+}^2 \vert V_+ \vert^2 \nonumber\\
& & \quad\quad\quad\quad \quad\quad\quad\quad \quad\quad
 +  \vert V_-'  \vert^2 - \Omega_{v-}^2 \vert V_- \vert^2 \Bigg].
 \end{eqnarray} 

An explicit computation gives
\begin{equation}
\frac{ \Omega_{v\pm}^2 }{ a^2 } = p^2 \pm \frac{\lambda}{f} \dot{\chi} p + \frac{ M_p^4 \left[  c_4 p^4 \pm c_3 p^3 + c_2 p^2 \pm c_1 p + c_0 \right]}{  M_p^2 p^2 \pm 2 g M_p^2 Q p + 2 g^2 Q^2 \left( M_p^2 + Q^2 \right) }
\label{om2vpm}
 \end{equation}
where $p=\frac{k}{a}$ is the physical momentum, and the coefficients $c_i$ are functions of background quantities. We note that
\begin{equation}
\Omega_{v+}^2 \left( p \right) = \Omega_{v-}^2 \left( - p \right).
\end{equation}
We also note that the numerator in (\ref{om2vpm}) is always positive. The exact expressions for $c_i$ are readily obtained, but they are not illuminating. We report here the leading expressions in the slow roll $\lambda \gg 1$ expansion:
\begin{eqnarray}
c_4 & = & H^2  \left[ \frac{2 g^2 Q^2}{H^2}  + {\rm O } \left( \frac{ 1 }{  \lambda } \right)  \right]  \nonumber\\
c_3 & = & H^3  \left[ \frac{ g^2 Q^2}{H^2} \left( 6 g \frac{Q}{H} + \frac{\lambda}{f} \frac{\dot{\chi}}{H} \right) + {\rm O } \left( \frac{ 1 }{  \lambda } \right)  \right]  \nonumber\\
c_2 & = & H^4  \left[ \frac{ g^2 Q^2}{H^2} \left( 3 +  4 g  \frac{\lambda}{f} \frac{\dot{\chi}}{H} + 8 g^2 \frac{Q^2}{H^2}  \right) + {\rm O } \left( \frac{ 1 }{  \lambda } \right)  \right]  \nonumber\\
c_1 & = & H^5  \left[ \frac{ g^4 Q^4}{H^4} \left( 4 g \frac{Q}{H} + 6  \frac{\lambda}{f} \frac{\dot{\chi}}{H} \right) + {\rm O } \left( \frac{ 1 }{  \lambda } \right)  \right]  \nonumber\\
c_0 & = & H^6  \left[ 
4 g^5 \frac{Q^5}{H^5} \frac{\lambda}{f} \frac{\dot{\chi}}{H}     + {\rm O } \left( \frac{ 1 }{  \lambda } \right)  \right]  \nonumber\\
\end{eqnarray}

Using (\ref{slow}), it is immediate to see that, in each of the above expression, the dominant term in the square parenthesis that we have written explicitly is a  ${\rm O } \left( 1 \right)$ coefficient. Moreover, the second term on the right hand side of  (\ref{om2vpm}) evaluates to $\pm p H \, {\rm O } \left( 1 \right) $. Therefore, at leading order in slow roll,
\begin{eqnarray}
&& \Omega_{v\pm}^2 \simeq k^2 \;\;\;,\;\;\; p \gg H \nonumber\\
&& \Omega_{v\pm}^2 \simeq \frac{f_A f_{\dot{\chi}}}{y^{3/4} } a^2 H^2 \;\;\;,\;\;\; p \ll H 
\label{omega-v}
\end{eqnarray}
in the deep sub-horizon and super-horizon regime, respectively. We conclude that the vector sector is stable.

We note that the super horizon limit  of (\ref{omega-v}) is precisely the slow roll expression of $a^2 m_g^2$ in the $m_g \gg H$ limit.  The two dynamical vector modes originate from the perturbations of the vector field, and the expression for the mass in this limit coincides with what found in  \cite{Dimastrogiovanni:2012st}.

\section{Scalar modes}
\label{sec:scalars}

After gauge fixing, the system of scalar perturbations comprises of $1$ mode from the inflaton, $\delta \chi$, $3$ from the SU(2) field, $Y, \delta Q, M$,  and $2$ from the metric, $\phi$ and $B$. Among,  these $6$ modes, $\delta \chi, \delta Q$ and $M$ are dynamical, while the other three modes are nondynamical; namely they enter in the quadratic action without time derivatives, and they can be integrated out (see the discussion at the start of the previous section). Integrating out the metric perturbations makes the algebra extremely involved; for this reason, in the study presented in the main text we make the approximation of setting $\phi = B = 0$ from the start. We then integrate out  $Y$, and we denote the resulting quadratic  action by  $ S_{2,{\rm scalar}}$. We also performed the full computation, which we present in Appendix \ref{app-dg-scalar}. 
We denote by $S_{2,{\rm scalar-full}}$ the quadratic action obtained by including all the modes and by integrating out the $3$ nondynamical ones. Both  $ S_{2,{\rm scalar}}$ and   $S_{2,{\rm scalar-full}}$ are functionals of the $3$ dynamical modes. 

We expand all entries in these actions in slow roll. As we show in Appendix  \ref{app-dg-scalar}, all entries of the  matrices of the two actions agree at the leading order in this expansion at all scales (namely, for all values of $H/p$), with a single exception. The exception is the  the $11$ coefficient of the frequency matrix $\Omega^2$, for which the agreement is excellent only up to   $H \la 10 p$. The discrepancy that takes place afterwards  is surely completely  irrelevant for the stability study that we perform here (as we will see, the instability, if present, manifests itself deeply inside the horizon).  Not surprisingly, the metric perturbations do not affect the stability of the background solution. Very likely, this disagreement has also no significant consequence for the power spectra that we show below, since it manifests itself only after the power has frozen (see the appendix for a detailed study).

We therefore  set  $\phi = B = 0$, and integrate out $Y$. We then define
\begin{eqnarray}
\delta \chi & \equiv & \frac{\Delta_1}{a} \nonumber\\
\delta Q & \equiv & \frac{\Delta_2}{\sqrt{2} a} \nonumber\\
\delta M & \equiv & \frac{g a Q \Delta_2 + \sqrt{k^2 + 2 g^2 a^2 Q^2} \, \Delta_3}{\sqrt{2} g k^2 a^2 Q}
\label{canonical}
\end{eqnarray}
In terms of the multiplet $\Delta = \left( \Delta_1 , \Delta_2 , \Delta_3 \right)^T$, the quadratic action for the scalar modes acquires, up to a total derivative, the form (\ref{formal}). We denote as $K_s$ and $\Omega_s^2$ the $3 \times 3$ matrices entering in this action. These matrices have the following entries
\begin{eqnarray}
\frac{K_{s,12}}{a} & = & \frac{g \lambda Q^2}{\sqrt{2} f } \nonumber\\
\frac{K_{s,13}}{a} & = & - \frac{g^2 \lambda Q^3}{\sqrt{2} f \sqrt{p^2+2 g^2Q^2}} \nonumber\\
\frac{K_{s,23}}{a} & = & 0
\label{Ks-dg0}
\end{eqnarray}
and
\begin{eqnarray}
\frac{\Omega_{s,11}^2}{a^2} & = & p^2 + \frac{g^2 \lambda^2 p^2 Q^4}{f^2 \left( p^2 + 2 g^2 Q^2 \right)} + \frac{g^2 Q^4}{M_p^2}
+ V_{,\chi\chi} \nonumber\\
& &  - 2H^2 + \frac{\dot{\chi}^2}{2 M_p^2} + \frac{\left( \dot{Q} + H Q \right)^2}{M_p^2} \nonumber\\
\frac{\Omega_{s,12}^2}{a^2} & = & \frac{3 g \lambda H Q^2}{\sqrt{2} f} + \frac{\sqrt{2} g \lambda Q \dot{Q}}{f} \nonumber\\
\frac{\Omega_{s,13}^2}{a^2} & = & - \frac{\sqrt{2} \lambda}{f} \Bigg[ \frac{g^2 H Q^3}{2 \sqrt{p^2 + 2 g^2 Q^2}} \nonumber\\
& & + \frac{2 p^4 + 3 g^2 p^2 Q^2 + 4 g^4 Q^4}{2 \left( p^2 + 2 g^2 Q^2 \right)^{3/2}} \left( \dot{Q} + H Q \right) \Bigg] \nonumber
\end{eqnarray}
\begin{eqnarray}
\frac{\Omega_{s,22}^2}{a^2} & = & p^2 + 4 g^2 Q^2 - \frac{g \lambda Q \dot{\chi}}{f} \nonumber\\
\frac{\Omega_{s,23}^2}{a^2} & = & - \sqrt{p^2 + 2 g^2 Q^2} \left( 2 g Q - \frac{\lambda}{f} \dot{\chi} \right) \nonumber\\
\frac{\Omega_{s,33}^2}{a^2} & = & p^2 + \frac{4 g^2 Q^2 \left( p^2 + g^2 Q^2 \right)}{p^2 + 2 g^2 Q^2} 
  - \frac{g \lambda p^2 Q \dot{\chi} }{f \left( p^2 + 2 g^2 Q^2 \right)}  \nonumber\\
& &  + \frac{6 g^2 p^2 \left( \dot{Q} + H Q \right)^2}{\left( p^2 + 2 g^2 Q^2 \right)^2} 
\label{Os-dg0}
\end{eqnarray}
We stress that these expressions are obtained by disregarding the scalar metric perturbations, but that they are otherwise exact.
We can also verify that the eigenvalues of the $\left\{ i , k \right\} = \left\{ 1 , 2 \right\}$ part of  $\Omega_{ij}^2$ are $\simeq a^2 m_g^2$ in the super horizon regime and for $m_g \gg H$, in agreement with  \cite{Dimastrogiovanni:2012st}.

We solve the theory specified by this action following the steps outlined in Subsection \ref{subsec:quantization}. We assume that, after inflation, only the inflaton field provides a sizable contribution to reheating (we note that the energy in $A_\mu$ is much smaller than the inflaton energy during inflation). In this limit, we have the curvature perturbation $\zeta \simeq - \frac{H}{\dot{\chi}} \delta \chi$, with the power spectrum
\begin{equation}
P_\zeta = \frac{H^2 \, {\cal P}_{11}}{\dot{\chi}^2}
\label{P-zeta}
\end{equation}
where ${\cal P}$ is given in (\ref{power}), with ${\cal Y}_i$ being the three fields on the left hand side of (\ref{canonical}).

\begin{figure}[ht!]
\centerline{
\includegraphics[width=0.35\textwidth,angle=-90]{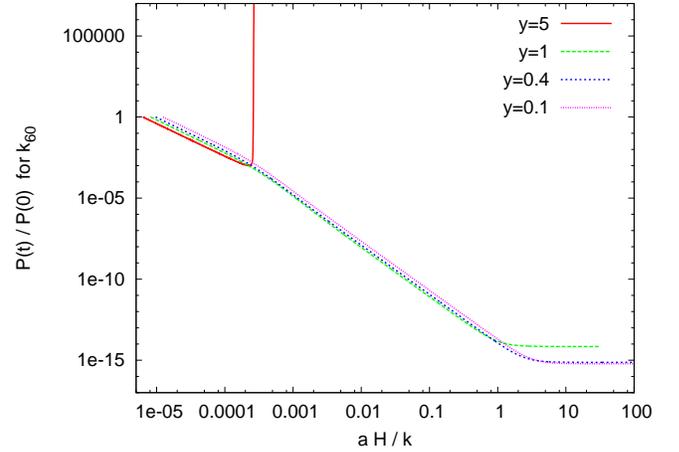}
}
\caption{Time evolution of the power (normalized to one at the initial time shown), for a mode that leaves the horizon $60$ e-folds before the end of inflation, and for the same background evolutions shown in Figure  \ref{fig:bck}. 
}
\label{fig:Pkt}
\end{figure}

In Figure \ref{fig:Pkt}, we then present the time evolution of ${\cal P}_{11}$ for a single mode (a given $k$) and for the same choices of background  parameters that we used in Figure \ref{fig:bck} for the background evolution. For definiteness, we considered in all cases the mode that leaves the horizon $60$ e-folds before the end of inflation, and we denoted the corresponding comoving momentum by $k_{60}$. We choose the initial time of the evolution such that the mode is deeply inside the horizon at the start, and the last term in (\ref{Omega}) is negligible. We observe that the choice $y=5$ leads to an instability of the linearized theory, while  the other three cases are stable, and are characterized by a power that freezes outside the horizon, as in the standard inflationary models.

The instability of the $y=5$ choice manifests itself at some  scale inside the horizon, namely $p_{\rm unstable } = k / a_{\rm unstable} \gg H$. Even the stable solutions show a different evolution at this scale (note in the Figure the small change in the slope of $P \left( t \right)$ for the stable cases at the same scale at which the $y=5$ solution becomes unstable). We can see this analytically, by considering the approximated expressions of  the scalar system inside the horizon. We will obtain the scale  $\Lambda H$ times a numerical factor, where
\begin{equation}
\label{lambd}
\Lambda \equiv \frac{  \sqrt{\lambda} }{ y^{1/4 } } \,  \frac{M_p}{f} \;\;\;,\;\;\; \Lambda \gg 1
\end{equation}
(in the numerical examples that we have shown, $\Lambda  = \sqrt{2} \, 10^3 \, y^{-1/4}$). Let us discuss the approximation in more details. First of all, using the slow roll conditions (\ref{slow}), we find  $g^2 Q^2 \simeq y^{-3/2} f_A^2 H^2 \ll \Lambda^2 H^2 $ in all cases that we have studied. Therefore,  we can disregard $g^2 Q^2$ in comparison to $p^2$ inside eqs. (\ref{Ks-dg0}) and  (\ref{Os-dg0}). From this, and from the slow roll approximations (\ref{slow}), we obtain, in the sub-horizon regime
\begin{eqnarray}
\frac{K_{s,12}}{a} & = & \frac{g \lambda Q^2}{\sqrt{2} f } = {\rm O } \left( \Lambda H \right) \nonumber\\
\frac{K_{s,13}}{a} & = &   {\rm O } \left( \Lambda H \, \frac{H}{p} \right) \ll K_{12,s}
\label{K-s-sub}
\end{eqnarray}
and
\begin{eqnarray}
& &  \frac{\Omega_{s,11}^2}{a^2}  \simeq  p^2 + \frac{g^2 \lambda^2 Q^4}{f^2 } = {\rm O } \left( \Lambda^2 H^2 \right)  \nonumber\\
& & \frac{\Omega_{s,12}^2}{a^2}  =  {\rm O } \left( \Lambda H^2 \right) \nonumber\\
& & \frac{\Omega_{s,13}^2}{a^2}  \simeq  - \frac{\sqrt{2} \lambda p}{f}    H Q 
 = {\rm O } \left( \Lambda H p \right)  \nonumber\\
& & \frac{\Omega_{s,22}^2}{a^2} , \frac{\Omega_{s,33}^2}{a^2}    =  p^2 +  {\rm O } \left(  H^2 \right) \nonumber\\
& & \frac{\Omega_{s,23}^2}{a^2}  =  {\rm O } \left(  H \, p  \right) \nonumber\\
\label{O-s-sub}
\end{eqnarray}

We want to solve the evolution equations (\ref{Delta-eom}) until $p \sim \Lambda H$. We can do so by rewriting them as
\begin{equation} 
\ddot{\cal D}_{ij} + \frac{K_{s,ik}}{a} \dot{\cal D}_{kj} + \frac{\Omega_{s,ik}^2}{a^2} {\cal D}_{kj} \simeq 0
\label{app-sca-eq}
\end{equation}
by inserting only the terms written explicitly in (\ref{K-s-sub}) and (\ref{O-s-sub}), and by treating these terms as constant. All these approximations amount in considering only terms  that contribute to the dynamics at $ {\rm O } \left( p^2 ,\, \Lambda H p ,\, \Lambda^2 H^2 \right)$.
We note for instance that $\dot{\cal D} = {\rm O } \left( p {\cal D} \right)$ in this regime, so that it is consistent to set $\Omega_{s,12}^2=0$ in the approximated equation, while retaining $K_{s,12}$ and the dominant term of $\Omega_{s,13}^2$. We also note that the fastest evolving coefficient in the matrices is $p$ that evolves on a ${\rm O } \left( H^{-1} \right)$ timescale; therefore time derivatives of the terms in 
 (\ref{K-s-sub}) and (\ref{O-s-sub}) introduce at most terms with an additional factor $H$, which are therefore suppressed in the 
  $p > \Lambda H$ and  $p \simeq \Lambda H$ regimes.

Performing these approximations, the system (\ref{app-sca-eq}) reduces to a set of linear second order equations with constant coefficients.
The equations split in three separate groups, one for the complex unknowns ${\cal D}_{i1}$ (with $i=1,2,3$), one for the  the complex unknowns ${\cal D}_{i2}$, and  one for the three complex unknowns  ${\cal D}_{i3}$. In each group we need to solve three second order differential equations, and therefore we have $6$ possible solutions; we note that the three groups have the identical set of equations, and they differ only in the initial conditions. Therefore the solutions are of the type
\begin{equation}
{\cal D} \simeq \sum_{a=1}^6 C_a \, {\rm e}^{\xi_a  t}
\end{equation}
where the matrices $C_a$ are integration constants (so to match the initial conditions) and $\xi_a$ are constant numbers. The system is unstable if any of the $\xi_a$ has a real and positive part. By solving the system, one can see that the only coefficient that can possibly be real is
\begin{eqnarray}
\xi & = & \Bigg[ - p^2 - \frac{3 g^2 \lambda^2 Q^4}{2 f^2}  \nonumber\\
& & \quad\quad + \frac{\lambda Q}{f} \sqrt{ 2 p^2 \left( H^2 + g^2 Q^2 \right) + \frac{9}{4} \frac{\lambda^2 g^4 Q^6}{f^2} } \Bigg]^{1/2}
\label{xi}
\end{eqnarray}
and therefore
\begin{equation}
{\rm stability} \; \Leftrightarrow \; p^2 > {\bar p}^2 \equiv
\frac{\lambda^2 Q^2}{f^2} \left( 2 H^2 - g^2 Q^2 \right)
\label{stability-k}\end{equation}
where, using the slow roll condition, 
\begin{equation}
     \frac{{\bar p}^2}{H^2} = \Lambda^2 \, y^{1/2}  \, f_A^2 f_H^2 \left( 2 - \frac{f_A^2}{y^{3/2}} \right)
     \end{equation}

If the expression in parenthesis is negative the background solution is stable. Otherwise, there is an instability at sufficiently large wavelengths. Therefore
\begin{equation}
{\rm stability} \; \Leftrightarrow \; y < \frac{\sin^{2/3} x }{2 \left( 1 + \cos x \right)^{4/3}} \,. 
\label{stability-param}
\end{equation}
For any fixed $y$, this condition is violated at sufficiently small $x$.  This provides an upper bound to the amount of inflation in the model (for any given choice of parameters). One needs to verify that this upper bound is comparable with the required amount of inflation.  For the evolutions shown in Figure \ref{fig:bck}, the inflaton field, at $60$ e-folds before the end of inflation, evaluates to $\chi \simeq 2.45 f$ for $y=5$, to  $\chi \simeq 2.20 f$ for $y=1$, to $\chi \simeq 2.00 f$ for $y=0.4$, and to  $\chi \simeq 1.56 f$ for $y=0.1$. Correspondingly, the rhs of the condition (\ref{stability-param}) evaluates to $\simeq 2.62$ for $y=5$, to $\simeq 1.43$ for $y=1$,  to    $\sim 0.95$ for $y=0.4$, and  to $\simeq 0.49$ for $y=0.1$.   The criterion (\ref{stability-param}) therefore indicates that the choice $y=5$ does not lead to a stable inflationary solution of $60$ e-folds (recall that $f$ and $\lambda$ are fixed to $10^{-2} M_p$ and to $500$, respectively, in this example), while the other choices do. This is in perfect agreement with Figure \ref{fig:Pkt}.
 
Eq. (\ref{stability-k}) confirms that the instability, if it exists, takes place at $p < {\bar p}=\Lambda H$ times an order one factor. For example, for the unstable $y=5$ choice, the criterion (\ref{stability-k}) gives an instability starting at ${\bar p} \simeq 2.67 \Lambda H$, corresponding to $a H/{\bar p} \simeq 0.00025$, in excellent agreement with the evolution seen in the Figure.

\begin{figure}[ht!]
\centerline{
\includegraphics[width=0.35\textwidth,angle=-90]{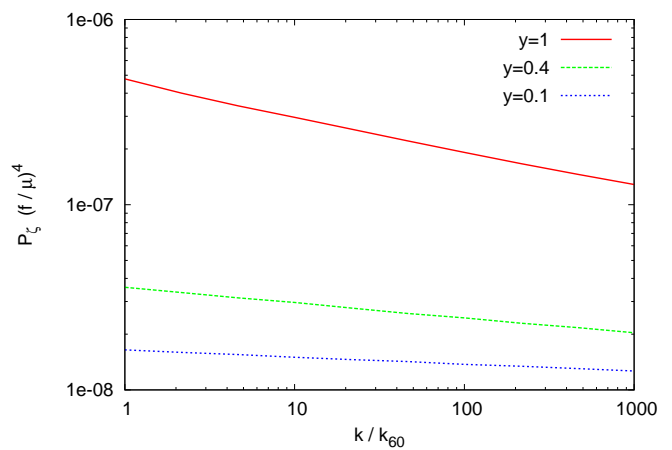}
}
\caption{Power spectra for $ \lambda = 500$ and three different values of $y$. The spectral index 
(defined as $P_\zeta \propto k^{n_s-1}$) is $n_s \simeq 0.81$ for $y=1$, 
 $n_s \simeq 0.92$ for $y=0.4$ and  $n_s \simeq 0.96$ for $y=0.1$. 
}
\label{fig:spe}
\end{figure}

Finally, let us verify that the instability is extremely fast. Let us assume that we have a background solution with $2 H^2 > g^2 Q^2 $. At any given moment,  modes with $p \geq {\bar p}$ have ${\rm Re } \xi = 0$, and are therefore stable. Moreover, the instability is also negligible at very large scales, given that  $\xi \rightarrow 0$ for $p \rightarrow 0$. However, for $p \la {\bar p}$, a quick study of eq. (\ref{xi}) shows that $\xi = {\rm O } \left(  \Lambda H \right)$ for $p$ smaller than, but parametrically equal to ${\bar p}$. (for $2H^2 \gg g^2 Q^2$, the maximum of $\xi$ is obtained for $p = {\bar p}/2$). This corresponds to a very short instability time $\sim \frac{1}{H \Lambda} \ll H^{-1}$. Therefore, each mode experiences a strong instability while still inside the horizon, and we conclude that the background solution is unstable whenever the condition (\ref{stability-param}) is violated. We conclude this Section by showing in  Figure \ref{fig:spe} the power spectrum $P_\zeta$ for the stable $y=0.1,0.4,1$ configurations, which we discuss in the next Section.

\section{Conclusions}
\label{sec:conclusions}

We performed a complete study of the linear order quantum fluctuations for the chromo-natural inflation model. We separated the metric, gauge and axion fluctuations into scalar, vector and tensor modes, verifying explicitly that they decouple at the linearized level. We computed their equations of motion and worked out the quantization of the system. The tensor sector consists of the two gravity wave polarizations plus two modes from the gauge field. The gauge mode of one helicity becomes tachyonic for some finite interval of time next to horizon crossing, sourcing one gravity wave helicity.  The vector sector consists of two dynamical modes that originate from the vector field, and that remain perturbatively small. The scalar sector contains three dynamical modes, one originating from the inflation and two from the gauge field. 

 We showed that, for some parameter choice, one of the eigenfrequencies of this system, that we denoted by $\xi$, can become  imaginary 
 inside the Hubble horizon, leading to a fast instability.  Let us compare this with a standard result in inflation. Specifically, let us  consider the gravity  wave amplitude $\delta g_{ij} = a^2 h_{ij}^{TT}$. The canonical variable $h_c \propto a h^{TT}$ obeys $h_c'' + \Omega^2 h_c = 0$ with the dispersion relation $ \Omega^2 = a^2 \left( p^2 - 2  H^2 \right)$. As a consequence $h_c \propto a$ outside the horizon, corresponding to a frozen amplitude of $h^{TT}$. Therefore, although the frequency of the canonical mode becomes tachyonic outside the horizon, its magnitude is not large enough to lead to a physical instability.  The situation is analogous for a test scalar field or for the scalar perturbations in the standard case. 
 
 For the case at hand, the eigenfrequency $\xi$ is given in eq. (\ref{xi}).   In this discussion, for illustrative purposes, let us approximate the full expression obtained in  eq. (\ref{xi}) with     $\xi^2 \sim p^2 -  c \Lambda^2  H^2$, where $\Lambda \gg 1$ is defined in (\ref{lambd}), and $c$ an order one factor. No instability appears if $c$ is negative. Otherwise, an instability takes place for modes of physical momentum $p = {\rm O } \left( \Lambda H \right)$. The timescale of the instability is $\vert \xi \vert^{-1} = {\rm O } \left( \frac{1}{\Lambda H} \right) \ll H^{-1}$,   (or, equivalently, the amplitude of the scalar modes grows proportionally to a large power of the scale factor) which indeed corresponds to a fast instability.     This heuristic discussion reproduces the results obtained from the precise form of $\xi$ (see Section \ref{sec:scalars} for the precise computation). Moreover, we also performed a fully numerical and exact study of the scalar perturbations that  confirms these analytical results.

  The parameters of the model can be chosen so that $\xi$ remains real inside the horizon (equivalently, $c < 0$ in the heuristic expression) so that the instability is avoided. This corresponds to choosing $m_g \equiv \sqrt{2} g Q > 2 H $, where $Q$ is the vev of the vector field, and $g$ the SU(2) coupling. Quite interestingly, the quantity $m_g$ coincides with the mass of the fluctuations of the vector field in the $m_g \gg H$ regime   \cite{Dimastrogiovanni:2012st}, and it approximates well this mass for $m_g \ga H$. Therefore, the inflationary solution is stable if and only if the vector field is sufficiently heavy. 
 
We have obtained the boundary of the stable region $m_g \gg H$ analyzed in  \cite{Dimastrogiovanni:2012st} and our formalism can be readily employed to study the phenomenology of the boundary region. This study  is beyond the purposes of the present work. For illustrative purposes, we have however computed the power spectrum of $\zeta$ for some sample choices of parameters. Specifically, Figure \ref{fig:spe} shows the power spectrum for an axion decay constant $f = 0.01 M_p$, for $\lambda = 500$ (we stress that 
the main motivation for the model is to provide inflation for a sub-Planckian axion decay constant, and that this can be obtained for sufficiently large $\lambda$), and for three choices of $y$. We note that, among   those shown in  Figure \ref{fig:spe}, only the power spectrum obtained for $y=0.1$ is sufficiently flat to meet the observational bounds  \cite{Komatsu:2010fb,Hinshaw:2012fq}.  We numerically found that the  spectral tilt is a decreasing function of $y$ in the  $0.1 < y < 1$  interval (equivalently, it is an increasing function of $m_g/H$ in this interval). This behavior can also be seen in the analytic relation given in   \cite{Dimastrogiovanni:2012st} in the large $m_g$ regime. Therefore, both the requirement of stability and of flatness of the power spectrum pose a lower bound on $m_g/H$.  

\begin{figure}[ht!]
\centerline{
\includegraphics[width=0.35\textwidth,angle=-90]{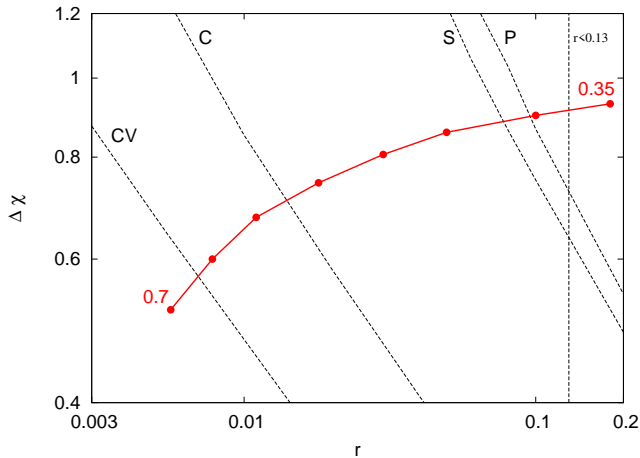}
}
\caption{Red/solid curve: Value of $\Delta \chi$ and $r$ obtained for  $f = 0.01 M_p$ and  $\lambda = 500$, and for different choices of $y$ in the $0.35 < y < 0.7$ interval (successive points along the curve denote $0.05$ increments in $y$); black/dotted vertical line:  $r<0.13$  bound from \cite{Hinshaw:2012fq};  the other black/dotted curves are the $1 \sigma$  detection lines for the Planck (P), SPIDER (S), CMB-Pol (C), and a cosmic-variance limited (CV) experiment. The signal needs to be above a line to be detectable at $ 1 \sigma$ by that experiment. These experimental forecasts are an approximate copy of the lines shown in Figure 2 of  \cite{Gluscevic:2010vv}.
}
\label{fig:r-dchi}
\end{figure}

On the other hand, the discussion around eq. (\ref{mgH}) leads to the conclusion that a  large $m_g/H$   leads to a detectable or ruled out gravity wave signal. The  enhanced signal is parity violating, which  results in nonvanishing  TB and EB correlations. A measure of the net handedness of the tensor modes is
\begin{equation}
\vert \Delta \chi \vert \equiv \Big\vert \frac{P_L - P_R}{P_L + P_R} \Big\vert
\end{equation}
where $P_{L/R}$ is the power spectrum of the left/right-helicity gravity wave modes. The corresponding observational bounds have been studied in \cite{Saito:2007kt,Gluscevic:2010vv}. In Figure \ref{fig:r-dchi} we compare the bounds presented in \cite{Gluscevic:2010vv} with the value of $r$ and $\Delta \chi$ obtained in this model, for our choice of $f = 0.01 M_p$ and  $\lambda = 500$, and for different values of $y$ in the $0.35 < y < 0.7$ range (the tensor power spectra are   obtained from the $11$ element of (\ref{power}) for the left and right helicity tensor sectors, following the procedure outlined in Subsection \ref{subsec:quantization}). Greater values of $y$ do not lead to a visible gravity wave signal even in a cosmic variance limited experiment, while lower values are ruled out by the current limit $r < 0.13$  \cite{Keisler:2011aw,Hinshaw:2012fq}.

We point out that all values of $y$ considered in this plot are actually ruled out because they lead to a too small value of $n_s$ (the largest value $n_s \simeq 0.93$ is obtained for $y = 0.35$, and $n_s$ then further decreases at larger $y$).  We stress that our choice of 
$f = 0.01 M_p$ and  $\lambda = 500$ is only dictated by the requirement of $f \ll M_p,\, \lambda \gg 1$, so that the coupling to the gauge field plays a relevant role for the inflation dynamics (which, in turns, is the initial purpose of the model \cite{Adshead:2012kp}). Obviously, these values can be changed, and our  phenomenology discussion has the only purpose of understanding what kind of limits can be imposed on the model. We expect that a viable region will exist at larger $f$ (and smaller $\lambda$), as the model becomes closer to a free inflaton model slowly rolling on a flat potential.

At the theoretical level, it is interesting to note   that the model can result in an observable  gravity wave signal in the near future,  even if the inflaton spans a range which is some order of magnitudes smaller than the Planck scale. This   evades the Lyth bound    \cite{Lyth:1996im}, that states that $r \ga 0.01$ is possible only for an excursion of the inflaton of ${\rm O } \left( M_p \right)$ during the last $\sim 60$ e-folds of inflation. We note that the model we have studied evades the bound because of the $\chi F {\tilde F}$ interaction.  Quite interestingly, the only other examples that we are aware of where $r > 0.01$ can be achieved even if the inflation evolution is orders of magnitude below the Planck scale~\footnote{We note that  inflationary potentials leading to $r \ga 0.01$ have been constructed where the inflationary range is smaller than, but still of ${\rm O } \left( M_p \right)$ \cite{BenDayan:2009kv,Hotchkiss:2011gz}.}   are those  studied in refs. \cite{sol6a,Barnaby:2012xt}, which are  characterized by the same pseudo-scalar interaction. As we already mentioned, ref.  \cite{sol6a} is also characterized by a $\sim f \ll M_p$ evolution due to the damping from gauge field production. In this case, an interesting parity-violating gravity wave signal can be generated   \cite{Sorbo:2011rz} from the produced gauge quanta, but one needs to evade the simultaneous generation of too many non-gaussian scalar density perturbations   \cite{Barnaby:2010vf}. Ref.  \cite{Sorbo:2011rz} achieves this by considering a large number of gauge fields (this reduces non-gaussianity by the central limit theorem) or by introducing a curvaton field.   Ref. \cite{Barnaby:2012xt}   shows that $r\ga 0.01$ can be obtained, and the non-gaussianity limit can be respected,  if the  rolling-scalar is not the inflaton. 

An important difference is that, however, in the mechanism of   \cite{sol6a,Barnaby:2012xt} the tensor modes are produced at the non-linear level by the vector fields produced by the rolling inflaton. For the model  \cite{Adshead:2012kp}, the production occurs already at the linear level, due to its mixing with the vector mode induced by the vector vev and the non-abelian structure ($g,Q \neq 0$). Quite likely, also in this model the vector modes $t_L$ can source a significant amount of  scalar density perturbations at the non-linear level. This may reduce $r$ from the level studied here, although   a too large $m_g$ will still likely be ruled out by the significant gravity wave production. It is  possible that, for regimes resulting in acceptable $r$, the sourced scalar modes will lead to interesting levels of non-gaussianity and primordial black holes as those obtained in \cite{Barnaby:2010vf,Barnaby:2011vw,Meerburg:2012id,Linde:2012bt}. It is also possible that, for some choice of parameters,  the vector field production will "self-regulate" (with a consequent decrease of $r$) due to its  backreaction on the inflationary dynamics, as visible in the background evolutions studied in  \cite{sol6a,Barnaby:2011qe} (a too large production may slow down the inflaton, and this may in turns decrease the vector production). All these interesting possibilities remain to be studied.

 \vspace{0.5cm}
{\bf Note added:} In the first version of this manuscript, we pointed out that (i) the inflationary solution of \cite{Adshead:2012kp} is stable if and only if $m_g> 2 H$, and that (ii)  the tensor-to-scalar ratio $r$ in this model is enhanced with respect to the case of a free inflaton, leading to violation of the Lyth's bound \cite{Lyth:1996im} for some choice of parameters. Both these claims are confirmed by the present analysis. The stability study is unchanged with respect to the first version. The study of the tensor modes, and the phenomenology considerations that follow from it,  have instead been updated, to include  the helicity violating terms in the tensor action that were erroneously missing in the first version. Such terms  result in a further increase of $r$ and in a helicity violating gravity wave signal. The relevance of these terms  was pointed out in  \cite{Adshead:2013qp}, that appeared on the archive between the first and the current version of this manuscript. Ref.  \cite{Adshead:2013qp} agrees with our limit   $m_g> 2 H$ for the stability of the inflationary solution, and our revised analysis    agrees with the effect of the  helicity violating effects found in     \cite{Adshead:2013qp}.

\section*{Acknowledgments}

MP acknowledges Nicola Bartolo and Sabino Matarrese for useful conversations. This work  was supported in part by DOE grant DE-FG02-94ER-40823 at the University of Minnesota.   MP would like to thank the University of Padova, and INFN, Sezione di Padova, for their friendly hospitality and for partial support during his sabbatical leave.

\appendix

\section{Including scalar metric perturbations}
\label{app-dg-scalar}

In Section \ref{sec:scalars} we studied the system of linear perturbations for the model, disregarding the perturbations of the metric. Here we summarize the results for the full system, and we confirm the accuracy of the approximation made in the main text.

We start from the full set of $6$ scalar perturbations (after gauge fixing), including the $2$ modes from the metric perturbations ($\phi$ and $B$) that we have (artificially) set to zero in the main text. Both these modes are non-dynamical, and we integrate them out, together with the other non-dynamical perturbation $Y$. We are left with three dynamical modes, and we ``rotate'' them as in eq. (\ref{canonical}) of the main text. In this way, we obtain the full quadratic action for the perturbations. It is formally of the type
\begin{eqnarray}
&&
S_{2,{\rm scalar-full}}  =  \frac{1}{2} \int d \tau d^3 k \Bigg[ \Delta^{' \dagger} C_{\rm s,f}  \Delta' 
\nonumber\\
& & \quad\quad \quad\quad 
+  \Delta^{' \dagger} K_{\rm s,f} \Delta - \Delta^\dagger K_{\rm s,f}  \Delta' - \Delta^\dagger \Omega_{\rm s,f}^2 \Delta \Bigg]
\label{formal-S2sf}
\end{eqnarray}

Namely, the modes  (\ref{canonical}) are not the exact canonical scalar variables. However, as we now show, they provide a very good approximation to the canonical modes in the slow roll regime. More in general, the matrices of the full action (\ref{formal-S2sf}) are extremely involved. We studied them in slow roll approximation. Specifically, using the slow roll approximation (\ref{slow}), we can cast all the elements of these matrices in the form
\begin{equation}
\frac{\sum_i c_i k^{p_{\alpha_i}} H^{\beta_i}}{\sum_i d_j k^{p_{\alpha_j}} H^{\beta_j}}
\end{equation}
where the coefficients $c_i$ and $d_j$  only depend on the parameters of the model, and on slowly evolving  background quantities.
We computed the leading order expression for these coefficients in the slow roll approximation. For example, we obtain
\begin{eqnarray}
\left( C_{\rm s,f} \right)_{11} & = & 1 + \frac{g^2 Q^4 \dot{\chi}^2 / M_p^4}{2 H^2 p^2 + 4 g^2 H^2 Q^2 \left[ 1 + {\rm O } \left( \frac{1}{\lambda} \right) \right]} \nonumber\\
& = & 1 + \frac{{\rm O } \left( \frac{f^2}{M_p^2 y^{3/2} \lambda^3} \right) H^4}{2 H^2 p^2 + {\rm O } \left( \frac{1}{y^{3/2}} \right) \left[ 1 + {\rm O } \left( \frac{1}{\lambda} \right) \right] H^4} \nonumber\\
\end{eqnarray}
This expression is extremely close to one, since, parametrically, the second term is $\lambda^{-3} \ll 1$ outside the horizon, and even more suppressed inside the horizon.       In fact, the kinetic matrix differs from the identity only up to slow roll suppressed quantities:
\begin{eqnarray}
\left( C_{\rm s,f} \right)_{12} & = & \frac{{\rm O } \left( \lambda^{1/2} \right) H^2}{ {\rm O } \left( y^{3/2} \lambda^3 \right) p^2 + {\rm O } \left( \lambda^3 \right) H^2} \nonumber\\
\left( C_{\rm s,f} \right)_{13} & = & \frac{{\rm O } \left( \lambda^{3/2} \right) H \sqrt{H^2 + {\rm O } \left( y^{3/2} \right) p^2}}{ {\rm O } \left( y^{3/2} \lambda^3 \right) p^2 + {\rm O } \left( \lambda^3 \right) H^2} \nonumber\\
\left( C_{\rm s,f} \right)_{22} & = & 1 +  \frac{{\rm O } \left( \lambda \right) H^2 }{ {\rm O } \left( y^{3/2} \lambda^3 \right) p^2 + {\rm O } \left( \lambda^3 \right) H^2} \nonumber\\
\left( C_{\rm s,f} \right)_{23} & = & \frac{{\rm O } \left( \lambda^{2} \right) H \sqrt{H^2 + {\rm O } \left( y^{3/2} \right) p^2}}{ {\rm O } \left( y^{3/2} \lambda^3 \right) p^2 + {\rm O } \left( \lambda^3 \right) H^2} \nonumber\\
\left( C_{\rm s,f} \right)_{33} & = & 1 +  \frac{{\rm O } \left( \lambda^2 \right) H^2 }{ {\rm O } \left( y^{3/2} \lambda^3 \right) p^2 + {\rm O } \left( \lambda^3 \right) H^2} \nonumber\\
\end{eqnarray} 
Therefore, up to very small slow roll corrections, the modes (\ref{canonical}) are also the canonical variable of the full scalar system.

Performing the same procedure on $K_{\rm s,f}$, we obtain
\begin{eqnarray}
\left( \frac{K_{\rm s,f}}{a} \right)_{12} & = & \frac{g \lambda Q^2}{\sqrt{2} f} \, \frac{p^2 + m_g^2 \left[ 1 + {\rm O } \left( \frac{1}{\lambda} \right) \right]}{p^2 + m_g^2 \left[ 1 + {\rm O } \left( \frac{1}{\lambda} \right) \right]} \nonumber\\
& = & \left( \frac{K_{\rm s}}{a} \right)_{12}  \left[  1 + {\rm O } \left( \frac{\lambda^{-1}}{ 1 + y^{3/2} \, p^2 / H^2 }  \right) \right]
\nonumber\\
\left( \frac{K_{\rm s,f}}{a} \right)_{13} & = & - \frac{g^2 \lambda Q^3}{\sqrt{2} f \sqrt{p^2+m_g^2}}
\, \frac{\left( p^2 + m_g^2 \right) \left[ 1 + {\rm O } \left(  \frac{1}{\lambda} \right) \right] }{p^2 + m_g^2 \left[ 1 + {\rm O } \left( \frac{1}{\lambda} \right) \right]}
 \nonumber\\
& = &  \left( \frac{K_{\rm s}}{a} \right)_{13} \left[ 1 + {\rm O } \left( \lambda^{-1} \right) \right] \nonumber\\
\left( \frac{K_{\rm s,f}}{a} \right)_{23}  & = & \frac{H^4}{\left( p^2+m_g^2 \right)^{3/2}}
\left[ {\rm O } \left( \frac{1}{y^{9/4} \lambda} \right) \frac{p^2}{H^2} + {\rm O } \left( \frac{1}{y^{15/4} \lambda} \right)  \right]
\nonumber\\
& \sim & {\rm O } \left( \frac{H}{\sqrt{p^2+H^2}} \frac{f}{M_p} \frac{1}{\lambda^{3/2}} \right) \times 
\left( \frac{K_{\rm s,f}}{a} \right)_{12} 
\end{eqnarray}
where  we recall that  $m_g^2 \equiv 2 g^2 Q^2$.  The $12$ and $13$ entries are in excellent agreement with those given in the main text. We recall that the  $13$ entry is much smaller than the $12$ entry inside the horizon, and it is negligible in the stability study. We note that, for the full system, the $23$ element is nonvanishing, while $\left( K_{\rm s} \right)_{23} = 0$. However, this element is strongly slow roll suppressed with respect to the other two, and completely negligible.

Proceeding in the same way (for brevity, we omit here the powers of $y$), we obtain
\begin{eqnarray}
\frac{\left( \Omega_{\rm s,f}^2 \right)_{12}   }{ \left( \Omega_{\rm s}^2 \right)_{12} } & = &   1 + \frac{
\sum_{i=0}^2 {\rm O } \left( H^{2i} p^{4-2 i} \right) }{ \left( p^2 + m_g^2 \right)^2 \lambda }  \nonumber\\
\frac{ \left( \Omega_{\rm s,f}^2 \right)_{13} }{    \left( \Omega_{\rm s}^2 \right)_{13} } & = & 
  1 + \frac{1}{\lambda} \frac{H^2}{p^2 + m_g^2} \frac{  \sum_{i=0}^2 {\rm O } \left( H^{2i} p^{4-2 i} \right)  }{  \sum_{i=0}^2 {\rm O } \left( H^{2i} p^{4-2 i} \right)  }    \nonumber\\
\frac{\left( \Omega_{\rm s,f}^2 \right)_{22} - k^2 }{\left( \Omega_{\rm s}^2 \right)_{22} - k^2 } & = & 1 + {\rm O } \left( \frac{1}{\lambda} \right) \nonumber\\
\frac{\left( \Omega_{\rm s,f}^2 \right)_{23}   }{ \left( \Omega_{\rm s}^2 \right)_{23} } & = & 
1 + \frac{  \sum_{i=0}^3 {\rm O } \left( H^{2i} p^{6-2 i} \right)  }{ \left( p^2 + m_g^2 \right)^3  \lambda }  \nonumber\\
\frac{\left( \Omega_{\rm s,f}^2 \right)_{33} - k^2 }{\left( \Omega_{\rm s}^2 \right)_{33} - k^2 } & = & 1 + 
\frac{  \sum_{i=0}^3 {\rm O } \left( H^{2i} p^{6-2 i} \right)  }{ \left( p^2 + m_g^2 \right) \lambda  \sum_{i=0}^2 {\rm O } \left( H^{2i} p^{4-2 i} \right)  }  \nonumber\\
\end{eqnarray}
and we see that, for all these entries, the expressions of $  \Omega_{\rm s,f}^2 $ and of $  \Omega_{\rm s}^2  $ agree at all scales (namely, for any value of $H/p$) up to subdominant ${\rm O } \left( \lambda^{-1} \right)$ corrections.  For the $11$ entry, we obtain
\begin{equation}
\frac{\left( \Omega_{\rm s,f}^2 \right)_{11} - k^2 }{\left( \Omega_{\rm s}^2 \right)_{11} - k^2 }  =
1 + \frac{H^2}{p^2 \lambda} \, \frac{{\rm O } \left( p^2 \right) + {\rm O } \left( H^2 \right)}{p^2+m_g^2}
\label{om11-compa}
\end{equation}
Also on this entry, the two matrices are in perfect agreement during the full sub-horizon regime.  However, while for the other entries the agreement continues also in the super-horizon regime, the $11$ entries differ from each other for $H^2 >  {\rm O } \left( \lambda p^2 \right)$.
By evaluating the coefficients in (\ref{om11-compa}), we found that the disagreement starts only at $H \ga 10 p$ for all the evolutions studied in the main text.

To conclude, all the matrix elements, up to one exception, of the system of scalar perturbations studied in the main text (in which we made the approximation of disregarding the scalar metric perturbations) agree at all scales (namely, for any value of $H/p$) with the corresponding entries of the full system up to subdominant terms in a slow-roll expansion. The single exception is the $11$ entry of $\Omega^2$, for which the agreement persists during the entire sub-horizon regime, at horizon crossing, and also up to $H \la 10 p$, but not further.  This proves that the stability study performed in the main text is valid also when the metric perturbations are included, given that the instability, when present, manifests itself deeply inside the horizon.  Most likely, this guarantees that also the power spectra shown in the main text are accurate, since the disagreement manifests itself only after the powers have frozen (see Figure \ref{fig:Pkt}).

\end{document}